\documentclass[
 superscriptaddress,
 preprint,
 amsmath,amssymb,
 aps,
floatfix,
]{revtex4-2}

\usepackage{graphicx}
\usepackage{dcolumn}
\usepackage{bm}

\begin{document}


\title{Multiple gravity laws for human mobility within cities}

\author{Oh-Hyun Kwon}
\affiliation{Department of Physics, Pohang University of Science and Technology, Pohang, Republic of Korea}

\author{Inho Hong}
\affiliation{Graduate School of Data Science, Chonnam National University, Gwangju, Republic of Korea}
\affiliation{Center for Humans and Machines, Max Planck Institute for Human Development, Berlin, Germany}

\author{Woo-Sung Jung}
\affiliation{Department of Physics, Pohang University of Science and Technology, Pohang, Republic of Korea}
\affiliation{Department of Industrial and Management Engineering, Pohang University of Science and Technology, Pohang, Republic of Korea}

\author{Hang-Hyun Jo}
\email{h2jo@catholic.ac.kr}
\affiliation{Department of Physics, The Catholic University of Korea, Bucheon, Republic of Korea}

\date{\today}

\begin{abstract}
The gravity model of human mobility has successfully described the deterrence of travels with distance in urban mobility patterns. While a broad spectrum of deterrence was found across different cities, yet it is not empirically clear if movement patterns in a single city could also have a spectrum of distance exponents denoting a varying deterrence depending on the origin and destination regions in the city. By analyzing the travel data in the twelve most populated cities of the United States of America, we empirically find that the distance exponent governing the deterrence of travels significantly varies within a city depending on the traffic volumes of the origin and destination regions. Despite the diverse traffic landscape of the cities analyzed, a common pattern is observed for the distance exponents; the exponent value tends to be higher between regions with larger traffic volumes, while it tends to be lower between regions with smaller traffic volumes. This indicates that our method indeed reveals the hidden diversity of gravity laws that would be overlooked otherwise.
\end{abstract}

\maketitle

\section{Introduction}

Mobility as the fabric of human societies has been studied for understanding the mechanism of movements~\cite{Gonzalez2008Understanding, Song2010Modelling,Song2010Limits, Simini2012Universal, Yan2017Universal, Alessandretti2020Scales, Schlapfer2021Universal}, diffusive processes~\cite{Balcan2009Multiscale, Brockmann2013Hidden}, and its association with socioeconomic attributes~\cite{Lee2017Morphology, Bassolas2019Hierarchical, Kraemer2020Mapping, Moro2021Mobility, Bokanyi2021Universal, Fan2023Diversity} at both individual and population levels. Taking advantage of rich data, several population-level mobility models were developed to describe travel patterns as a function of geographical factors, e.g., population distribution and travel distance. The gravity model~\cite{Zipf1946P1, Erlander1990Gravitya, Jung2008Gravity}, intervening opportunities model~\cite{Stouffer1940Intervening, Liu2020Universal}, and radiation model~\cite{Simini2012Universal,Ren2014Predicting,Kang2015Generalized,Alis2021Generalized} have been the leading concepts for the population-level mobility models~\cite{Barbosa2018Human}.

Among such mobility models, the gravity model has remained the representative population-level model, capturing the traffic in a simple form resembling Newton's gravity law; the traffic volume from one region to the other is proportional to the product of population sizes of origin and destination regions and inversely proportional to the distance between those regions~\cite{Zipf1946P1, Erlander1990Gravitya}. Precisely, its generalization is written as follows:
\begin{align}
\label{eq:gravity}
	T_{ij} = G \frac{m_i m_j}{r_{ij}^\gamma},
\end{align}
where $T_{ij}$ is the traffic volume between the origin region $i$ and the destination region $j$, and $r_{ij}$ is the geographical distance between regions $i$ and $j$. $m_i$ and $m_j$ are their population sizes or traffic volumes, while $G$ is a coefficient. The distance exponent $\gamma$ is to denote the deterrence of mobility on distance, and it usually takes a value ranging from 0.5 to 3~\cite{Barthelemy2011Spatial}. Note that the deterrence function of the gravity model has been observed in multiple forms, such as an exponential function~\cite{Liang2013Unraveling}, a combination of the exponential and power-law forms~\cite{Barbosa2018Human}, and the Hill function~\cite{Goh2012Modification}, besides the common power-law form. The gravity model has been applied to systems of various spatial interactions, including urban mobility~\cite{Lee2015Relating,Hong2016Application, Mazzoli2019Field, Li2021Gravity,Simini2021Deep,Ribeiro2023Mathematical}, intercity mobility~\cite{Jung2008Gravity, Liu2014Uncovering, Wang2019Gravity}, and general inter-regional interactions~\cite{Bhattacharya2008International, Krings2009Urban, Pan2012World, Palchykov2014Inferring, Lee2014Matchmaker, PrietoCuriel2018Gravity, Park2018Generalized, Kim2019Measuring}. 

For most applications of the gravity model, the entire set of data for a given geographical unit, e.g., cities or countries, has been analyzed to result in a single distance exponent, effectively ignoring the heterogeneities within the unit. Such heterogeneities can be accounted for by grouping the regions within the unit and then by separately estimating the distance exponent for each pair of origin and destination groups. This possibility of having multiple values of the distance exponent within the same unit has recently been suggested in a theoretical study considering heterogeneous population landscapes~\cite{Hong2019Gravity}. 

In our work, we test the validity of such ``multiple'' gravity laws on the commuting data in the twelve largest cities of the United States of America (USA) only using their traffic volumes, not the population data. For each city, we first divide urban areas into 10 groups by their traffic volumes and estimate the value of the distance exponent between each pair of groups. We find the varying patterns of the distance exponent for all cities analyzed and their commonality across different cities. Then, we compare the accountability of the multiple gravity laws to that of the conventional gravity model concerning only a single distance exponent for the entire city. We also discuss the origin of the multiple exponents in terms of the core-periphery structure of the city and travel costs. Our findings suggest that the gravity model on urban landscapes could be fine-tuned to incorporate the broad spectrum of urban movements for better understanding, estimation, and prediction.

\section{Data and method}

We analyze the commuting dataset processed from the LEHD Origin-Destination Employment Statistics (LODES) data in 2018~\cite{LODES}, where LEHD stands for the Longitudinal Employer-Household Dynamics project of the United States Census Bureau. The LODES data is a census survey connecting homes (origins) and workplaces (destinations) at a census block group (CBG) level. Here, census blocks are the smallest geographical unit for sampling the data, and a CBG consists of clusters of blocks, typically containing a few thousand people~\cite{GARM}. We use the Python module \texttt{geopandas} to derive geometric centroids of CBGs from geographical boundaries of CBGs in 2018~\cite{CBG}. The LODES data includes the number of trips between such CBGs.

In our work, we choose the twelve most populated Metropolitan Statistical Areas (MSAs) in the USA, which we call \textit{cities} hereafter. For each city, we divide the entire city into $1$~km~$\times$~$1$~km square \textit{cells} within the MSA boundary in 2018~\cite{MSA} on the Universal Transverse Mercator (UTM) coordinate system. Then each cell may contain several CBGs; the CBGs whose centroids are located in the same cell are merged to represent the traffic volume of the cell. Cells containing no CBGs' centroids are ignored for the analysis. Table~\ref{tab:data} shows the number of cells and the total number of trips between cells for each city.

\begin{table}[ht]
\centering
\caption{Twelve most populated USA cities for the analysis with information on the number of cells and the number of trips within each city.}
\begin{tabular}{c|c c}
\hline
City & Number of cells & Number of trips\\
\hline
New York City & 23,642    & 35,217,720\\
Los Angeles & 6,324     & 18,822,540\\
Chicago     & 12,777    & 15,185,472\\
Dallas      & 12,976   & 11,158,472\\
Houston     & 9,488    & 9,222,488\\
Washington, D.C.& 10,875 & 9,454,492\\
Miami       & 3,957     & 8,323,512\\
Philadelphia& 10,162     & 9,939,204\\
Atlanta     & 14,691    & 8,383,040\\
Phoenix     & 5,951     & 7,028,940\\
Boston      & 8,077     & 8,237,372\\
San Francisco & 3,225     & 6,530,276\\
\hline
\end{tabular}
\label{tab:data}
\end{table}

\begin{figure}[th!]
    \includegraphics[width=0.9\linewidth]{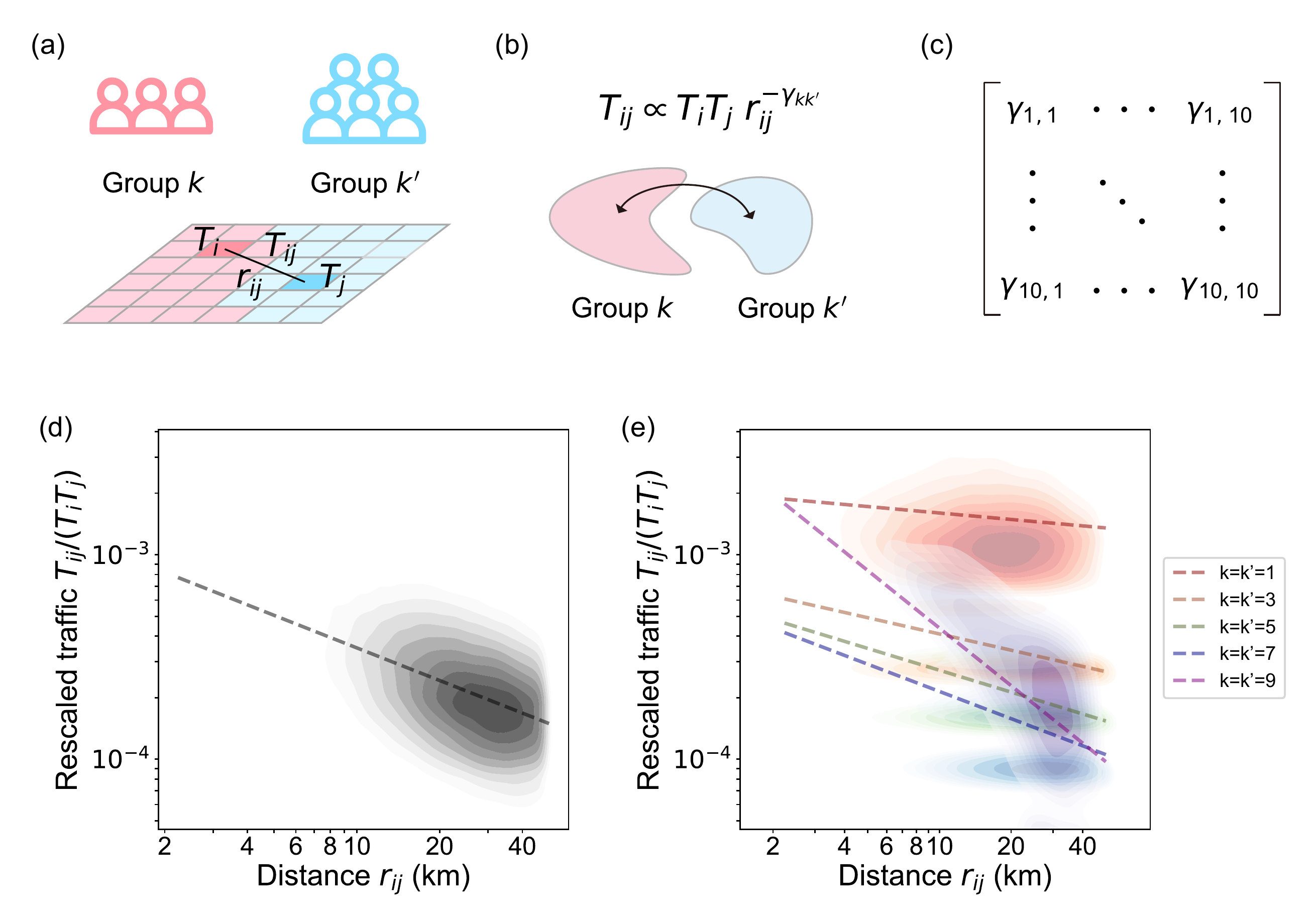}
    \caption{\textbf{Data analysis framework (a--c) and its application to the case of Chicago (d,e).} \textbf{(a)} For a given city, cells in a square grid are grouped into $10$ groups according to their traffic volumes $T_i$ in Eq.~\eqref{eq:Ti_define}. The traffic volume $T_{ij}$ and geographic distance $r_{ij}$ between a cell $i$ of group $k$ (red) and a cell $j$ of group $k'$ (blue) are identified. \textbf{(b)} Then, the distance exponent $\gamma_{kk'}$ is estimated using the gravity model in Eq.~\eqref{eq:fitting}. \textbf{(c)} Estimated values of the distance exponent, i.e., $\gamma_{kk'}$ for $k,k'\in\{1,\ldots,10\}$, form the exponent matrix $\Gamma$. \textbf{(d)} Conventional estimation of the distance exponent using the whole set of data for Chicago, resulting in a single distance exponent $\gamma_{\rm s}\approx 0.53$ (dashed line) in Eq.~\eqref{eq:fitting_single}. \textbf{(e)} Empirical confirmation of multiple gravity laws within Chicago with different values of $\gamma_{kk'}$ for some cases with $k=k'$ (dashed lines).}
    \label{fig:method}
\end{figure}

We describe the data analysis framework [see also Fig.~\ref{fig:method}(a--c)]. Each city has $N$ cells, and the numbers of trips, $T_{i\to j}$, from the $i$th cell to the $j$th cell for the pair of $i,j\in\{1,\ldots,N\}$ are given. Note that $T_{i\to j}\neq T_{j\to i}$ in general, since the commuting dataset only describes one-way trips from homes to workplaces. To illustrate the bidirectional traffic flow, we symmetrize the traffic volume between two cells as follows:
\begin{align}
\label{eq:symmetry}
    T_{ij}\equiv T_{i\to j}+T_{j\to i}.
\end{align}
Note that $T_{ii}=2T_{i\to i}$ by definition. The total number of trips for the $i$th cell is obtained by summing $T_{ij}$ over all $j$s:
\begin{align}
    T_i\equiv \sum_{j=1}^N T_{ij}.
    \label{eq:Ti_define}
\end{align}

The conventional gravity model assumes that the rescaled traffic volume between cells $i$ and $j$, defined as $T_{ij}/(T_iT_j)$, decays with the geographical distance between those cells, $r_{ij}$, as
\begin{align}
    \frac{T_{ij}}{T_iT_j}\propto r_{ij}^{-\gamma_{\rm s}} \ \textrm{for} \ i,j\in\{1,\ldots,N\},
    \label{eq:fitting_single}
\end{align}
where $\gamma_{\rm s}$ is the distance exponent for the entire set of pairs of cells in the city. Here we have rescaled $T_{ij}$ by the multiplication of traffic volumes of cells $i$ and $j$, i.e., $T_iT_j$, not by their populations as mentioned for Eq.~\eqref{eq:gravity}. We estimate the value of $\gamma_{\rm s}$ by means of the ordinary least squares linear regression using the equation: $\log(T_{ij}/T_iT_j)=A-\gamma_{\rm s}\log(r_{ij})+\epsilon$ with a constant $A$ and an error term $\epsilon$. The fitting range of $r_{ij}$ is limited to that showing the scaling behavior. Throughout the paper, we set the lower and upper bounds of the fitting range to $1$ km and $50$ km, respectively. We also calculate the $R^2$ value to quantify the quality of the fitting. For the case of Chicago we estimate $\gamma_{\rm s}\approx 0.53$ ($R^2\approx 0.17$) [Fig.~\ref{fig:method}(d)]. 

To investigate the possible variation of the gravity law within the city, we sort the cells according to their traffic volumes, $T_i$, and then sequentially group them into $10$ groups with an equal size of $N/10$. These groups are denoted by $G_k$ for $k=1,\ldots,10$, in ascending order of traffic volumes. The group of $k=1$ is for cells with the smallest traffic volumes, while the group of $k=10$ is for cells with the largest traffic volumes. Here we have grouped cells with respect to their traffic volumes in accordance with the previous theoretical work~\cite{Hong2019Gravity} that has shown the possibility of multiple gravity laws. We remark that there are alternative grouping methods, e.g., in Refs.~\cite{Bassolas2019Hierarchical, PrietoCuriel2021Heartbeat}. Now we estimate the distance exponent for each pair of groups, say $k$ and $k'$, assuming the following functional form:
\begin{align}
    \frac{T_{ij}}{T_iT_j}\propto r_{ij}^{-\gamma_{kk'}}\ \textrm{for}\ i\in G_k\ \textrm{and}\ j\in G_{k'}.
    \label{eq:fitting}
\end{align}
This functional form leads to the equation for the linear regression as $\log(T_{ij}/T_iT_j)=A-\gamma_{kk'}\log(r_{ij})+\epsilon$. Using the estimated values of $\gamma_{kk'}$ for all possible pairs of $k,k'\in\{1,\ldots,10\}$, we obtain the exponent matrix $\Gamma=[\gamma_{kk'}]$, as depicted in Fig.~\ref{fig:method}(c). Note that $\gamma_{kk'}=\gamma_{k'k}$ as we have symmetrized the traffic volumes between cells such that $T_{ij}=T_{ji}$, leaving us with $55$ distinct values of $\gamma_{kk'}$. In the case of Chicago, different scaling behaviors are observed with different values of $\gamma_{kk'}$. The results for several cases with $k=k'$ are shown in Fig.~\ref{fig:method}(e); e.g., $\gamma_{1,1}\approx 0.10$ ($R^2\approx 0.014$) and $\gamma_{10,10}\approx 0.94$ ($R^2\approx 0.37$) among others. Our data analysis framework can be applied to any mobility datasets as long as both $T_{ij}$ and $r_{ij}$ are available.

\section{Results}

\begin{figure}[t]
    \includegraphics[width=0.7\linewidth]{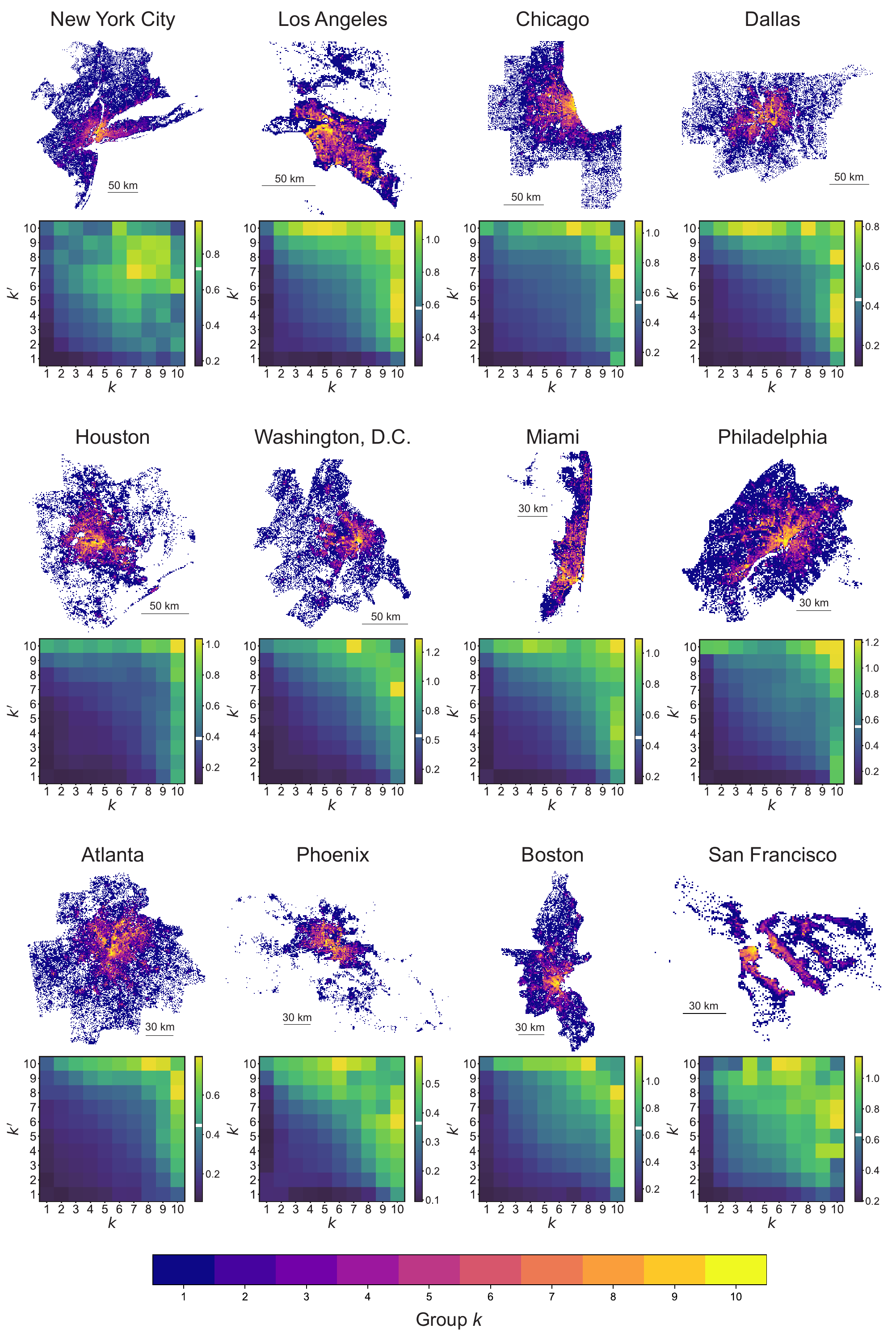}
    \caption{{\bf Traffic landscapes and exponent matrices for the twelve most populated cities in the USA.} Each panel consists of the traffic landscape of the city on the map (top) and the exponent matrix derived from the traffic volumes between cells in the city (bottom). In the landscapes, each cell is colored according to the group it belongs to, following the color bar at the bottom of the figure. A higher value of $k$ indicates the larger traffic volume of the group. For comparison, we mark the estimated distance exponent using the whole data of the city by a white bar in the color bar for exponent values.}
    \label{fig:exponent_matrix}
\end{figure}

\begin{figure}[th]
    \includegraphics[width=0.9\linewidth]{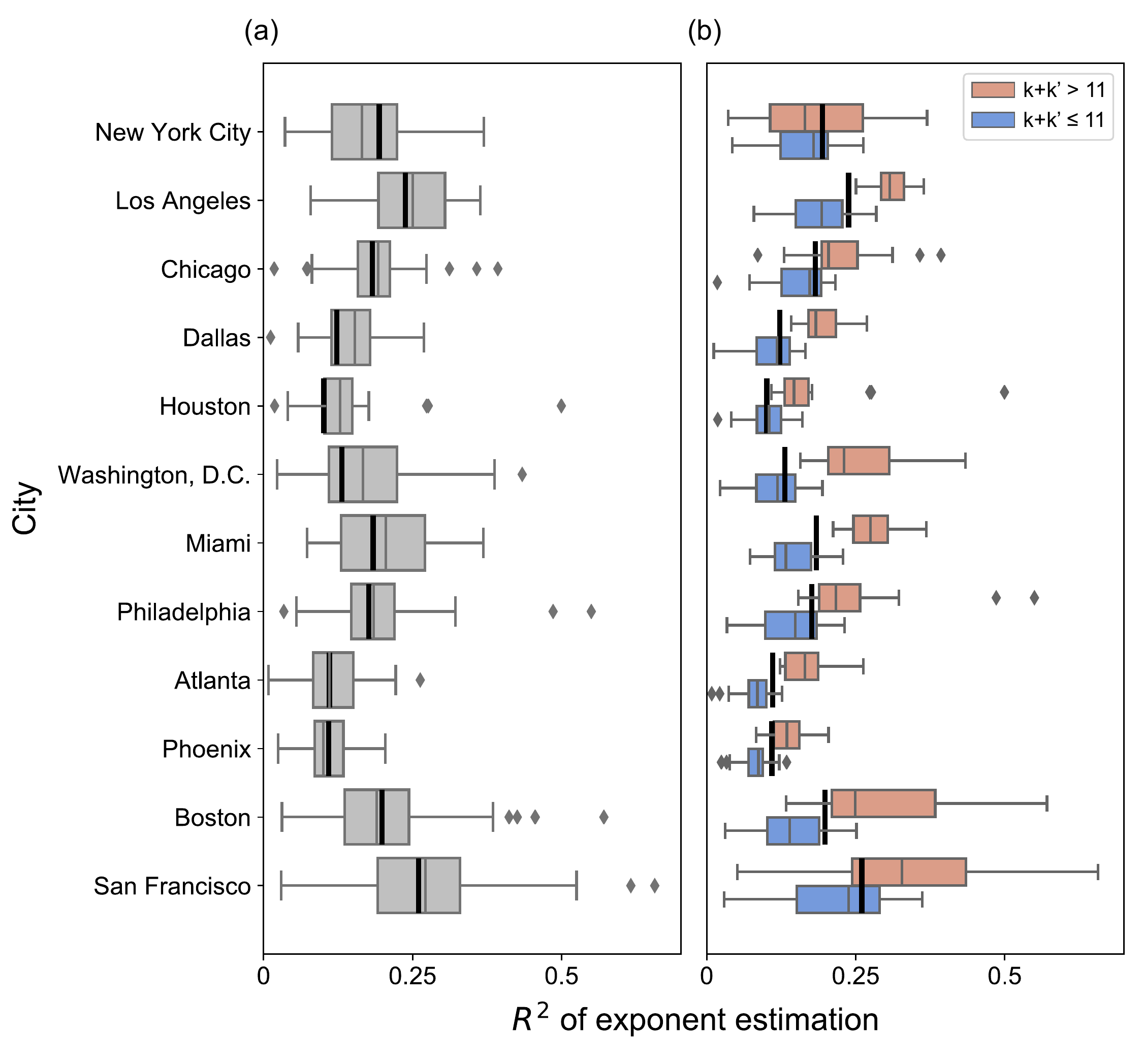}
    \caption{{\bf $R^2$ values in the distance exponent estimation for the twelve most populated cities in the USA.} \textbf{(a)} The box-and-whisker plot for each city describes a distribution of $R^2$ values for $55$ distance exponents in the exponent matrix, $\Gamma=[\gamma_{kk'}]$. The black thick vertical line indicates the $R^2$ value for the distance exponent, $\gamma_{\rm s}$, when using the whole set of data for each city. \textbf{(b)} The same results as in \textbf{(a)} are presented but by separating the cases with $k+k'>11$ (pink plots) from the other cases with $k+k'\leq 11$ (blue plots).}
    \label{fig:distribution}
\end{figure}

We apply the data analysis framework described in the previous Section to the twelve most populated cities in the USA as listed in Table~\ref{tab:data}. We first observe in Fig.~\ref{fig:exponent_matrix} that cells with the largest traffic volumes are concentrated at one or more centers of the cities. For example, cities such as Dallas and Houston have a single center, while cities like Los Angeles apparently have more than one center and those centers are distributed over the cities. On the other hand, cells with smaller traffic volumes are scattered over the cities.

The exponent matrices visualized in Fig.~\ref{fig:exponent_matrix} show that for each city the distance exponent $\gamma_{kk'}$ has various values according to the group indices $k$ and $k'$, which clearly evidence the multiple gravity laws within the city. For comparison, we mark as a white bar the value of the distance exponent, $\gamma_{\rm s}$, i.e., when using the whole set of pairs of cells in the city, in the color bar for the exponent values in Fig.~\ref{fig:exponent_matrix}. The finding of various values of $\gamma_{kk'}$ is consistent with our expectation that various scaling behaviors can be observed within the same city, depending on the populations or traffic volumes of both origin and destination cells~\cite{Hong2019Gravity}. Thus, our method can indeed reveal the hidden diversity of gravity laws that would be overlooked otherwise. In addition, one can say that the distance exponent $\gamma_{\rm s}$ might show an average behavior of the multiple gravity laws.

Interestingly, we find a common pattern in the exponent matrices of different cities, despite the huge diversity of the population, geographical constraints, and travel patterns of those cities; e.g., one can see a variety of geographical constraints such as sea, lakes, mountains, and/or neighboring cities in Fig.~\ref{fig:exponent_matrix}. To be precise, the value of the distance exponent $\gamma_{kk'}$ tends to be higher between groups of larger traffic volumes. The pairs of groups with large traffic volume and small traffic volume also tend to show higher values of $\gamma_{kk'}$ than those between groups with small traffic volumes. A possible explanation for such observations will be discussed later.

We examine the quality of the distance exponent estimation in terms of $R^2$ values. In Fig.~\ref{fig:distribution}(a), we show the $R^2$ value distribution for $\gamma_{kk'}$ [Eq.~\eqref{eq:fitting}] of each city in a box-and-whisker plot. The distribution is compared to the $R^2$ value for $\gamma_{\rm s}$ [Eq.~\eqref{eq:fitting_single}] of the same city, depicted as a black thick vertical line. We observe that the $R^2$ value for $\gamma_{\rm s}$ is not always better or worse than those for $\gamma_{kk'}$. To look at more details, we group $55$ exponent values into two subsets of roughly the same number of pairs; one subset includes $25$ cases with $k+k'> 11$ (i.e., pairs for relatively large traffic volumes), and the other subset is for $30$ cases with $k+k'\leq 11$ (i.e., pairs for relatively small traffic volumes). As shown in Fig.~\ref{fig:distribution}(b), it turns out that the $R^2$ values for $\gamma_{kk'}$ with $k+k'> 11$ tend to be much larger than that for $\gamma_{\rm s}$ in most cities, while the $R^2$ values for $\gamma_{kk'}$ with $k+k'\leq 11$ show the opposite tendency. The larger $R^2$ values for cases with larger $k$ and/or $k'$ indicate that multiple gravity laws can explain the mobility pattern for high-traffic regions better than the conventional gravity model characterized by a single distance exponent. On the other hand, the smaller $R^2$ values for cases with smaller $k$ and/or $k'$ might be due to the fact that cells with relatively small traffic volumes are scattered over the cities and mostly located at the periphery.

\begin{figure}[t]
    \includegraphics[width=0.35\linewidth]{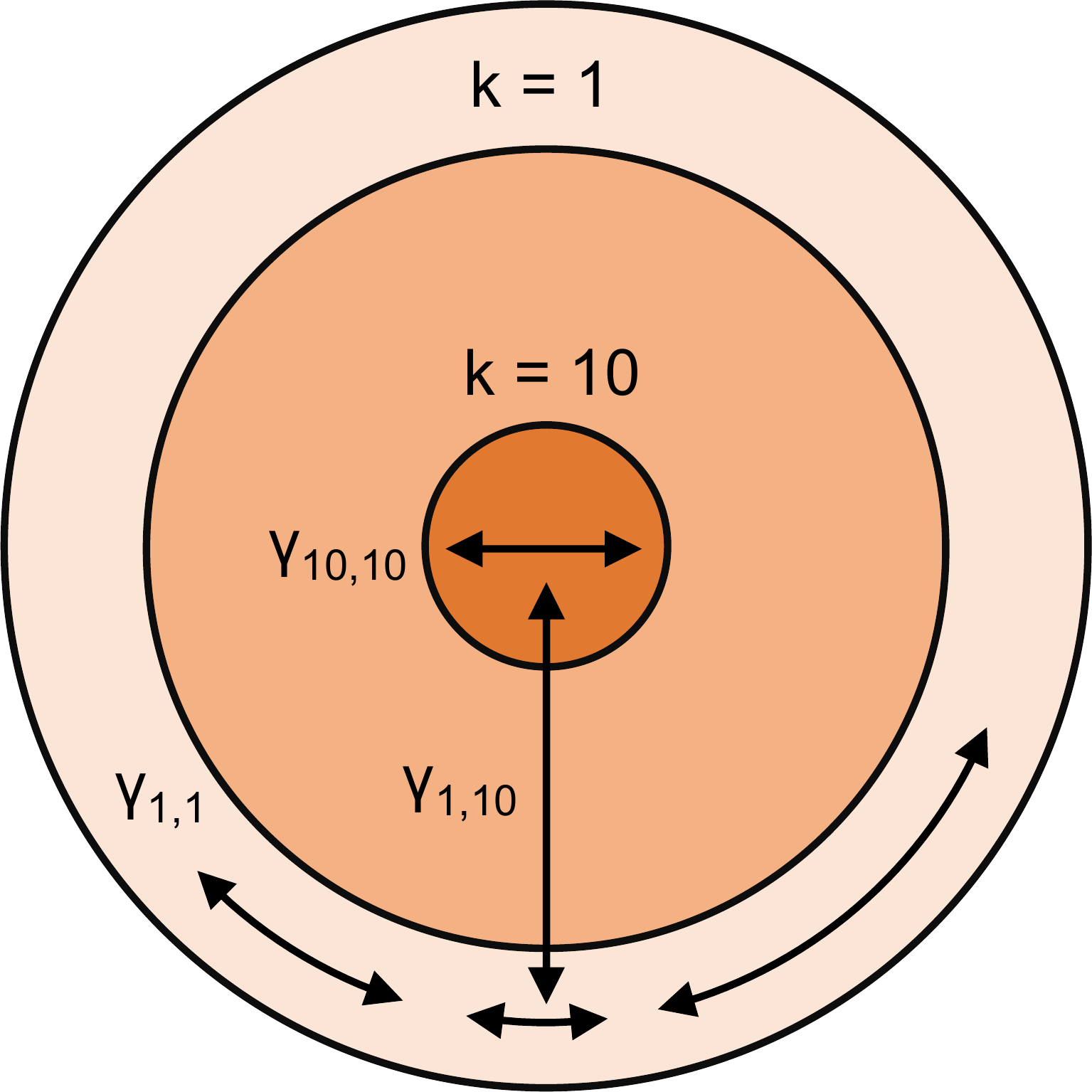}
    \caption{{\bf Schematic diagram for multiple gravity laws in a centralized urban landscape.} The dark shaded circle at the center and the bright shaded ring at the periphery denote groups of $k=1$ and $k=10$, respectively. Distance exponents $\gamma_{10,10}$, $\gamma_{1,1}$, and $\gamma_{1,10}$ characterize the distance dependence of traffic volumes between or within groups. Arrows visualize some possible trips between or within groups.}
    \label{fig:core-periphery}
\end{figure}

In order to understand the observed common pattern in the exponent matrices, we consider two main factors, namely, the traffic landscape and the travel cost. We find that the effects due to the traffic landscape seem to explain the observed exponent matrices to some extent, which can also be argued in terms of the travel cost. As evident in the traffic landscapes of Fig.~\ref{fig:exponent_matrix}, all the cities analyzed might be considered to have a so-called ``core-periphery'' structure whether the number of central areas or centers is one or more~\cite{Louf2013Modeling}. We remark that in contrast to cells with large traffic volumes, comprising the center(s), cells with small traffic volumes are scattered over the cities; see a schematic diagram for the core-periphery structure in Fig.~\ref{fig:core-periphery}. It implies that travel distances between cells in the periphery tend to be more diverse than those between cells in the center, thus possibly weakening the distance dependence of traffic volumes between cells in the periphery. Indeed, as shown in Fig.~\ref{fig:exponent_matrix}, the distance exponent $\gamma_{kk'}$ tends to have smaller values (e.g., $\lesssim 0.2$) for smaller $k$ and $k'$, while $\gamma_{kk'}$ has an overall higher value when either $k$ or $k'$ gets larger. 

\begin{figure}[t]
    \includegraphics[width=0.9\linewidth]{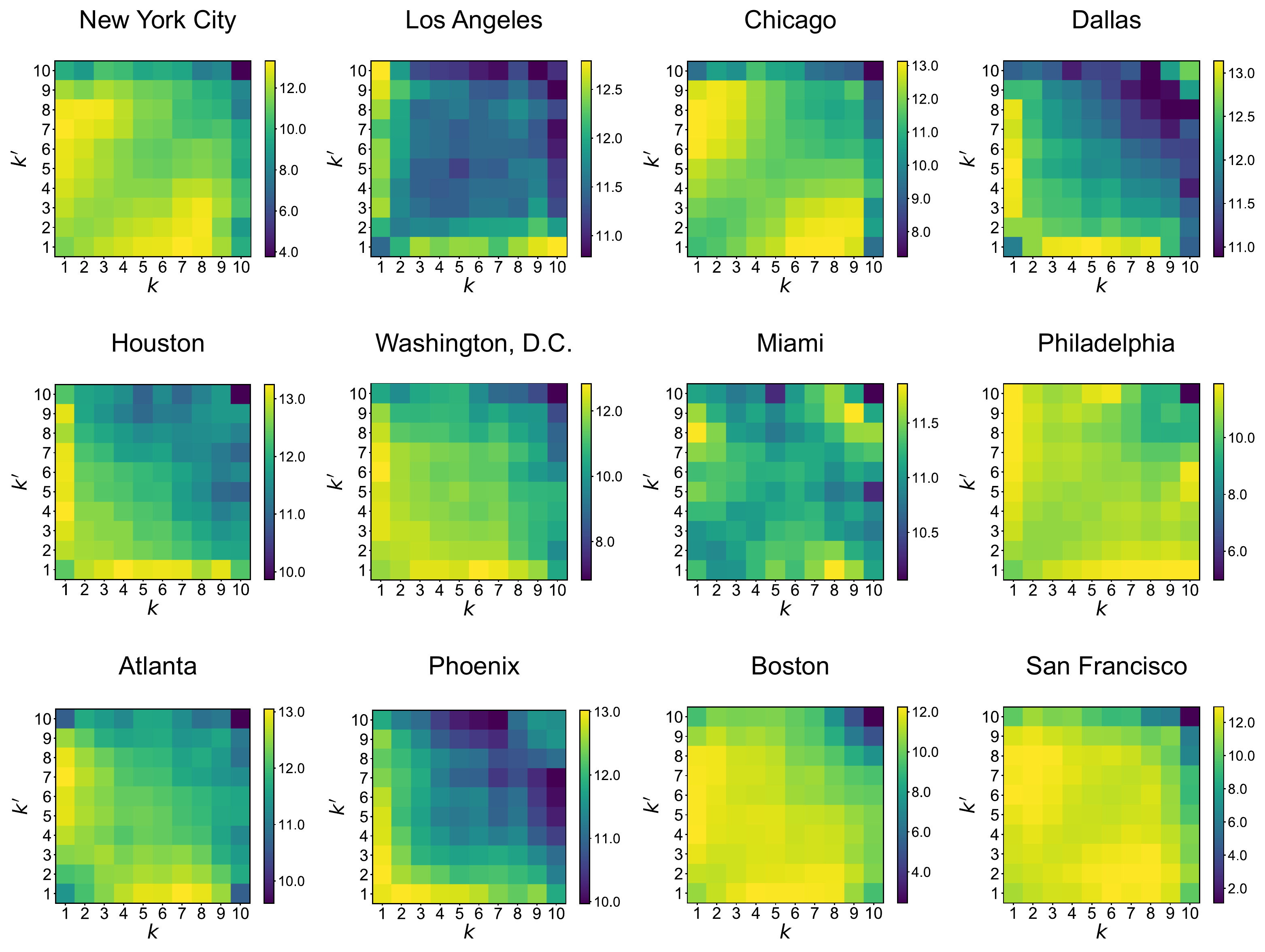}
    \caption{{\bf Standard deviations of travel distances between groups [Eq.~\eqref{eq:std_dist}] for the twelve most populated cities in the USA.} The color bar for the distance is in km. Note that cities have different ranges of distance in the color bar. Standard deviations of travel distances between groups of relatively small traffic volumes are overall larger than those of relatively large traffic volumes.}
    \label{fig:std_distance}
\end{figure}

\begin{figure}[t]
    \includegraphics[width=0.9\linewidth]{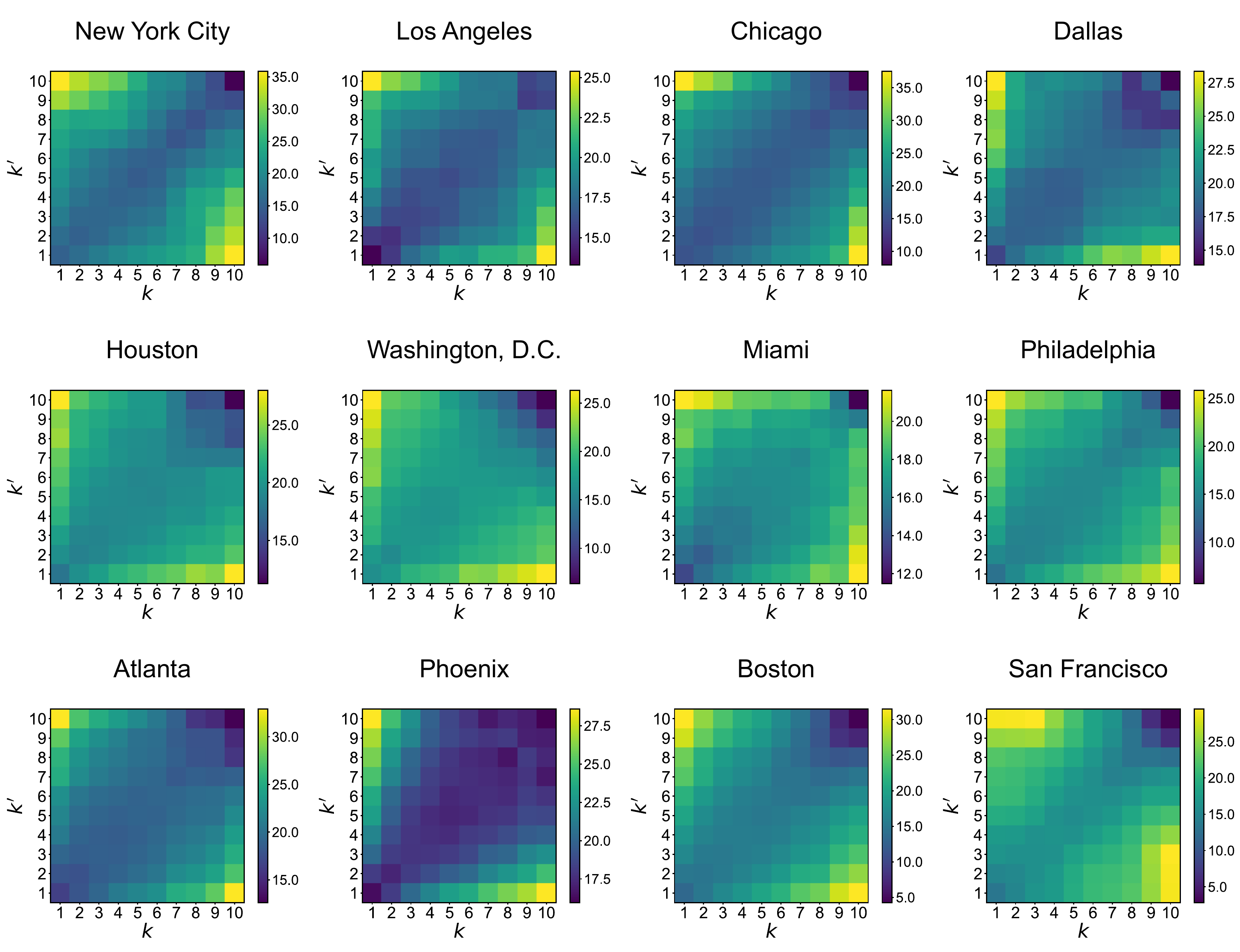}
    \caption{{\bf Average travel distances between groups [Eq.~\eqref{eq:avg_dist}] for the twelve most populated cities in the USA.} The color bar for the distance is in km. Note that cities have different ranges of distance in the color bar. Average travel distances between groups tend to be the largest between groups with the largest traffic volumes and the smallest traffic volumes than other pairs.}
    \label{fig:average_distance}
\end{figure}

Just because the possible travel distances are diverse does not necessarily mean that the real travel distances are diverse. To examine this issue, we calculate the standard deviation of travel distances between cells in two groups of $k$ and $k'$ which is defined as
\begin{equation}
    \sigma_{kk'} \equiv \left[\frac{\sum_{i\in G_k, j\in G_{k'}}T_{ij}\left(r_{ij}-\langle r\rangle_{kk'}\right)^2}{\sum_{i\in G_k, j\in G_{k'}}{T_{ij}}}\right]^{1/2},
    \label{eq:std_dist}
\end{equation}
where $\langle r\rangle_{kk'}$ is an average travel distance between cells in two groups of $k$ and $k'$:
\begin{equation}
    \langle r\rangle_{kk'} \equiv \frac{\sum_{i\in G_k, j\in G_{k'}}{T_{ij}r_{ij}}}{\sum_{i\in G_k, j\in G_{k'}}{T_{ij}}}.
    \label{eq:avg_dist}
\end{equation}
As shown in Fig.~\ref{fig:std_distance}, $\sigma_{kk'}$ tends to have higher values for smaller $k$ and/or $k'$ in all cities but Miami. This implies that the diversity of the travel distances between groups seems to be anti-correlated with the distance exponent values, as expected. As for the average travel distance $\langle r\rangle_{kk'}$, we find in Fig.~\ref{fig:average_distance} that the values of $\langle r\rangle_{kk'}$ for small $k$ and $k'$ are overall comparable to those for large $k$ and $k'$, both of which are shorter than those between groups of small traffic volumes and large traffic volumes. This tendency is denoted by long and short arrows in Fig.~\ref{fig:core-periphery}. Considering the fact that the dataset analyzed is for commuting between homes and workplaces, people living in the periphery do not travel so far from their homes, while people going to work to the center from the periphery (or the other way around) need to travel farther than others.

Next, we argue the effect of the travel cost for understanding the observed exponent matrices. The larger value of the distance exponent implies the stronger effect of the distance on the traffic volume, e.g., due to the higher travel cost per distance traveled. Here the travel cost can be measured in terms of elapsed time or transportation cost. In this sense, the larger values of the distance exponent within central areas or between central and peripheral areas could be due to the higher travel cost per distance. On the other hand, the smaller values of the distance exponent between peripheral areas could be partly due to the lower travel cost per distance. These arguments might be the case considering various factors in the central areas such as congestion, traffic signs, and speed limits that tend to increase the travel cost~\cite{Colak2016Understanding}. On the other hand, travels between peripheral areas tend to suffer from such factors less often, e.g., by taking highways. Yet our explanation is speculative at most, calling for more detailed empirical analyses in the future.

\section{Discussion}

In summary, we have devised the data analysis framework for urban mobility patterns and applied it to the dataset of the twelve most populated cities in the USA. We have found that the intra-city mobility patterns can be successfully characterized by multiple gravity laws, which means that the distance exponent value depends on the traffic volumes of the origin and destination regions within the same city. These findings are in contrast to the conventional gravity model characterized by a single distance exponent for a given dataset or area of interest. The common pattern in the distance matrices of different cities is observed, implying some common mechanisms behind such observations. 

The dataset we have analyzed has some limits. First, it includes only trips for commuting, but not other kinds of mobility such as shopping or tourism. Second, the dataset does not provide detailed information on the travel trajectory and cost, hampering further analysis to study the mechanisms for the multiple gravity laws.

In particular, information on travel trajectories within city centers and peripheral areas as well as between centers and peripheral areas must be relevant to understanding microscopic mechanisms behind the observed common pattern in exponent matrices. For example, with such information, one could study in more detail the impact of traffic conditions along trajectories between regions on their traffic volumes. Also, different geographical features, such as mountains and lakes, among cities might affect the travel trajectories in such cities, hence help us better understand the variation in our empirical findings. This approach will be complementary to the typical framework of gravity models only considering Euclidean distances between regions in the city.

Our findings suggest that the variation of distance exponent values can be used as an indicator to measure the appropriate dispersion of travel costs, as city centers with high traffic volumes tend to have large values of the exponent. For example, the difference in the exponent values before and after introducing new public transportation, e.g., subway or high-speed train, may be used to infer whether the new transportation has improved or redistributed the travel costs across the city.

Finally, we discuss possible future works. To investigate the mechanisms for the multiple gravity laws within the cities, one can study mathematical models considering the heterogeneous core-periphery structure of urban population and/or different travel costs depending on the mode of travel, etc. Based on the understanding of the mechanisms, one is expected to enhance the prediction and optimization of the mobility pattern within the city.


\begin{acknowledgments}
OHK and WSJ were supported by the National Research Foundation of Korea (NRF) grant funded by the Korea government (MSIT) (No. 2021R1F1A1063030).
IH was supported by the National Research Foundation of Korea (NRF) grant funded by the Korea government (MSIT) (RS-2023-00242528).
HHJ acknowledges financial support by the National Research Foundation of Korea (NRF) grant funded by the Korea government (MSIT) (No. 2022R1A2C1007358) and by The Catholic University of Korea, Research Fund, 2023.
\end{acknowledgments}

\bibliography{h2jo-papers}

\end{document}